 \documentclass[final,3p,times]{elsarticle}

\usepackage{graphicx} 
\usepackage{tabularx}
\usepackage{amsmath}
\usepackage{amssymb}
\usepackage{subfigure}
\usepackage{subfigure}
\usepackage{comment}
\usepackage{color}

\usepackage{multirow}

\usepackage{amssymb,amsmath,bm}

\definecolor{lightgray}{gray}{0.75}

\usepackage{fancyhdr}

\usepackage{lipsum}
\usepackage[dvipsnames]{xcolor}

\usepackage{hyperref}

\usepackage{cleveref}

\newcommand\myshade{85}
\colorlet{mylinkcolor}{violet}
\colorlet{mycitecolor}{YellowOrange}
\colorlet{myurlcolor}{Aquamarine}

\hypersetup{
  colorlinks=true,
  linkcolor  = mylinkcolor!\myshade!black,
  citecolor  = mycitecolor!\myshade!black,
  urlcolor   = myurlcolor!\myshade!black
}

\journal{CMA report}

\begin{document}

\begin{frontmatter}

\title{ A positivity-preserving conservative Semi-Lagrangian Multi-moment Global Transport Model on the Cubed Sphere}

\author[cnwp]{Jie Tang}
\author[xjtu]{Chungang Chen }
\author[cnwp]{Xueshun Shen}
\author[titech]{Feng Xiao}
\author[cnwp]{Xingliang Li \corref{cor}} 

\address[cnwp]{Center of Numerical Weather Prediction of NMC, China Meteorological Administration,  46 Zhongguancun South St., Beijing 100081, China }
\address[xjtu]{State Key Laboratory for Strength and Vibration of Mechanical Structures \& School of Human Settlement and Civil Engineering, Xi'an Jiaotong University, 28 Xianning West Road, Xifan, Shaanxi, 710049, China}
\address[titech]{ Department of Mechanical Engineering, Tokyo Institute of Technology, Tokyo 226-8502, Japan }

\cortext[cor]{Corresponding Address: Center of Numerical Weather Prediction, China Meteorological Administration,  46 Zhongguancun South St., Beijing 100081, China. Email address: lixliang@cma.gov.cn}

\begin{abstract}
A positivity-preserving conservative semi-Lagrangian transport model by multi-moment finite volume method has been developed on the cubed-sphere grid. In this paper, two kinds of moments, i.e. point values (PV moment) at cell boundaries and volume integrated average (VIA) value, are defined within a single cell. The PV moment is updated by a conventional semi-Lagrangian method, while the VIA moment is cast by the flux form formulation that assures the exact numerical conservation. Different from the spatial approximation used in CSL2 (conservative semi-Lagrangian scheme with second order polynomial function) scheme, a monotonic rational function which can effectively remove non-physical oscillations and preserve the shape, is reconstructed in a single cell by the PV moment and VIA moment. The resulting scheme is inherently conservative and can allow a CFL number larger than one. Moreover, the scheme uses only one cell for spatial reconstruction, which is very easy for practical implementation. The proposed model is evaluated by several widely used benchmark tests on cubed-sphere geometry. Numerical results show that the proposed transport model can effectively remove unphysical oscillations compared with the CSL2 scheme and preserve the numerical non-negativity, and it has the potential to transport the tracers accurately in real atmospheric model. 
\end{abstract}


\end{frontmatter}

\clearpage

\section{Introduction}
Global advection transport describes the motion of various tracers, which is a basic process in atmospheric dynamics. The numerical result of advection transport is significant in developing general circulation models (GCMs). Traditional latitude-longitude grid is very easy for application but has singularities at poles. Moreover, its nonuniform grid system would seriously affect computational efficiency. To address these issues, quasi-uniform grid systems without singularities or with weak singularities, such as the cubed-sphere grid, yin-yang grid and icosahedral grid, are becoming more and more popular in developing global transport model. Among them, cubed-sphere grid is much more attractive due to its computational merits, such as locally structured grid, local mass conservation and quasi-uniform grid. Some transport models based on the cubed-sphere grid can be seen in\cite{Chen2008, Chen2011, Guo2014,Guo2016,Lauritzen2010,Nair2005,Norman2018,Tang2018}. In this paper, we consider the cubed-sphere with gnomonic projection for our transport model.

Semi-Lagrangian method\cite{Staniforth1991} is widely used in transport model for it permit a large CFL number without reducing accuracy. But the traditional semi-Lagrangian method has a serious shortcoming, that is, it is not mass conservative.  To address this problem, many efforts have been made to develop the conservative semi-Lagrangian scheme. Nakamura et al.\cite{Nakamura2001} proposed such a scheme based on their previous Constrained Interpolation Profile (CIP) method\cite{Yabe1991}, and they called it CIP-CSL. In their method, the point values at cell boundaries and the cell-averaged value are used to construct the piecewise interpolation profile. The point values are updated by the semi-Lagrangian approach, while the cell average or volume average values are calculated by the flux-form formulation. The semi-Lagrangian approach permits a large CFL number, and the flux-form formulation of updating cell averaged values makes the scheme inherently conservative in terms of cell averaged values. Xiao and Yabe\cite{Xiao2001} introduced a slope limiter in CIP-CSL scheme to suppress oscillations around discontinuities, but the stencil for spatial reconstruction extended, from one cell to three cells. Instead of the cubic polynomial function used in CIP-CSL2, Xiao et al.\cite{Xiao2002} used a rational interpolation as a substitution, they called it CSLR, which used only one cell as stencil to construct interpolation function and remove non-physical oscillations simultaneously, but this scheme can’t completely preserve positivity. In this paper, we make some modification on the CSLR scheme to get a non-negative scheme and extend it to the cubed-sphere grid to develop a global transport model.

The paper is organized as follows. In section 2, we will review the 1D algorithm of CSLR scheme its modification. In section 3, we extend this formula to the cubed-sphere grid. Section 4 presents several kind of benchmark tests to evaluate the performance of the proposed global transport model. A brief summary is given in section 5.

\section{CSLR methods in one dimensional case}

\subsection{Spatial reconstruction}

\begin{figure}[htbp]
  \begin{center}
  \includegraphics[width=0.4\textwidth]{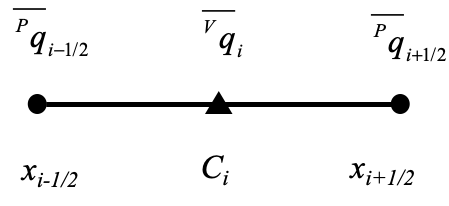}
  \end{center}
  \caption{Illustration of moments in one dimension..}\label{illustration1d}
  \end{figure}

To reconstruct the spatial profile, two kinds of moments are introduced in each cell, as illustrated in Fig. \ref{illustration1d}, PV moments at cell boundaries and VIA moment in $C_i$($i=1,2,\dots,N$)  are defined as

\noindent{\textbullet \quad The PV moments}
\begin{equation}
  \overline{^{P} q}_{i \pm 1 / 2}(t)=q\left(x_{i \pm 1 / 2}, t\right)
\end{equation}

\noindent{\textbullet \quad The VIA moment}
\begin{equation}
  \overline{^V q}_{i}(t)=\frac{1}{\Delta x} \int_{x_{i-1 / 2}}^{x_{i+1 / 2}} q(x, t) d x
\end{equation}
\noindent{where $q(x,t)$ is the transport quantity, $\Delta x$ is the grid spacing.}

Using these three moments, a rational function can be reconstructed in cell $i$

\begin{equation}
  R_{i}(x)=\frac{a_{i}+2 b_{i}\left(x-x_{i-1 / 2}\right)+\beta_{i} b_{i}\left(x-x_{i-1 / 2}\right)^{2}}{\left[1+\beta_{i}\left(x-x_{i-1 / 2}\right)\right]^{2}}
  \label{ri}
\end{equation}
\noindent{and the coefficients are determined by}

\begin{equation}
  R_{i}\left(x_{i-1 / 2}\right)=\overline{^P q}_{i-1 / 2}
  \label{ri:left}  
\end{equation}

\begin{equation}
  R_{i}\left(x_{i+1 / 2}\right)=\overline{^P q} _{i+1 / 2}
  \label{ri:right}
\end{equation}

\begin{equation}
  \frac{1}{\Delta x} \int_{x_{i-1 / 2}}^{x_{i+1 / 2}} R_{i}(x) d x=\overline{^v q}_i
  \label{ri:via}
\end{equation}

An alternative rational function can be expressed as

\begin{equation}
  R_{i}(x)=\frac{a_{i}+2 b_{i}\left(x-x_{i-1 / 2}\right)+\left(3 c_{i}+\beta_{i} b_{i}\right)\left(x-x_{i-1 / 2}\right)^{2}+2 \beta_{i} c_{i}\left(x-x_{i-1 / 2}\right)^{3}}{\left[1+\beta_{i}\left(x-x_{i-1 / 2}\right)\right]^{2}} \label{ri:new}
\end{equation}

\noindent{$\beta_i$  is the same as in Eq. \eqref{ri}. Using Eq. \eqref{ri:left} \eqref{ri:right} and \eqref{ri:via}, other coefficients can be determined. Details can be found in \cite{Xiao2001}.}

\subsection{Moments updating}

After getting the piecewise spatial reconstruction on the entire domain, the updating procedure can be conducted. Consider the following one-dimensional transport equation

\begin{equation}
  \frac{\partial q}{\partial t}+\frac{\partial(u q)}{\partial x}=0
  \label{transport equation}
\end{equation}

\noindent{\textbullet \quad Updating the PV moments:}

The PV moments are updated by the traditional semi-Lagrangian approach. Rewrite Eq. \eqref{transport equation} in an advection form

\begin{equation}
  \frac{\partial q}{\partial t}+u \frac{\partial q}{\partial x}=-q \frac{\partial u}{\partial x}
  \label{advection form}
\end{equation}
it can be viewed as an advection equation plus a source term $-q \frac{\partial u}{\partial x}$. The advection part is calculated by the semi-Lagragian concept

\begin{equation}
  \widetilde{^Pq}_{q_{i-1 / 2}}^{n+1}=R_{I}^{n}\left(x_{i p}\right)
  \label{updatapv}
\end{equation}

\noindent{where $x_{ip}$ is the departure point at previous time step $t=n\Delta t$ corresponding to the arrival point $x_{i-1/2}$ at next time step $t=(n+1)\Delta t$, the subscript $I$ is the index of the cell which contains the departure point $x_{ip}$, and departure point is simply calculated by }

\begin{equation}
  x_{i p}=x_{i-1 / 2}-\frac{u_{i-1 / 2}+u\left(x_{i p}^{1}\right)}{2} \Delta t
\end{equation}

\noindent {where $u\left( x_{ip}^1 \right)$ is the velocity at predict point $x_{ip}^1=x_{i-1/2}-u_{i-1/2}\Delta t$ . In general, $x_{ip}^{1}$  would not be identical with the point at cell interface, and the velocity at predict point  $x_{ip}^{1}$ is calculated by linear interpolation using known velocity at two interface of the cell which contains $x_{ip}^{1}$ . On the cubed-sphere grid, if predict point   is outside of patch boundary, the departure point is calculated by}

\begin{equation}
  x_{i p}=\left\{\begin{array}{ll}
    x_{-1 / 2}-u_{-1 / 2}\left(\Delta t-\left|\left(x_{i-1 / 2}-x_{-1 / 2}\right) / u_{i-1 / 2}\right|\right) & \text { if } x_{i p}^{1}<x_{-1 / 2} \\
    x_{N+1 / 2}-u_{N+1 / 2}\left(\Delta t-\left|\left(x_{N+1 / 2}-x_{i-1 / 2}\right) / u_{i-1 / 2}\right|\right) & \text { if } x_{i p}^{1}>x_{N+1 / 2}
    \end{array}\right.
\end{equation}

The ‘source term’ in Eq. \eqref{advection form} is simply approximated by

\begin{equation}
  \overline{^P q} _{i-1 / 2}^{n+1}=\widetilde{^P q}_{i-1 / 2}^{n+1}-\frac{\Delta t}{\Delta x_{i-1}+\Delta x_{i}}\widetilde{^P q} _{i-1 / 2}^{n+1}\left(u_{i+1 / 2}-u_{i-3 / 2}\right)
  \label{cal:source term}
\end{equation}

\noindent{\textbullet \quad	Updating the VIA moment:}

The VIA moment is updated by the flux-form concept

\begin{equation}
  \overline{^V q}_{i}^{n+1}=\overline{^V q}_{i}^{n}-\left(g_{i+1 / 2}-g_{i-1 / 2}\right) / \Delta x_{i}
  \label{updatevia}
\end{equation}

\noindent{where $g_{i+1/2}$ is the flux of $q$  going through the boundary $x_{i+1/2}$ during $[n \Delta t,(n+1) \Delta t]$, which is calculated by analytically integrating the interpolation function along the trajectory of $x_{i+1/2}$ }

\begin{equation}
  g_{i+1 / 2}=\int_{t^{n}}^{t^{n+1}} u_{i+1 / 2} q\left(x_{i+1 / 2}, t\right) d t
\end{equation}

\subsection{Modifications for positivity preserving and reducing overshoots}

For a problem with a lower boundary $q_{\text{min}}=0$  and upper boundary $q_{\text{max}}$ . The point values calculated by Eq.\eqref{cal:source term} may produce negative values and values exceed $q_{\text{max}}$ . An easy and effective modification for the PV moments is used

\begin{equation}
  \overline{^P q}_{i-1 / 2}^{n+1}=\min \left[\max \left(0, \overline{^P q}_{i-1 / 2}^{n+1}\right), q_{\max }\right]
  \label{modification0}
\end{equation}

\begin{figure}[htbp]
\begin{center}
\includegraphics[width=0.4\textwidth]{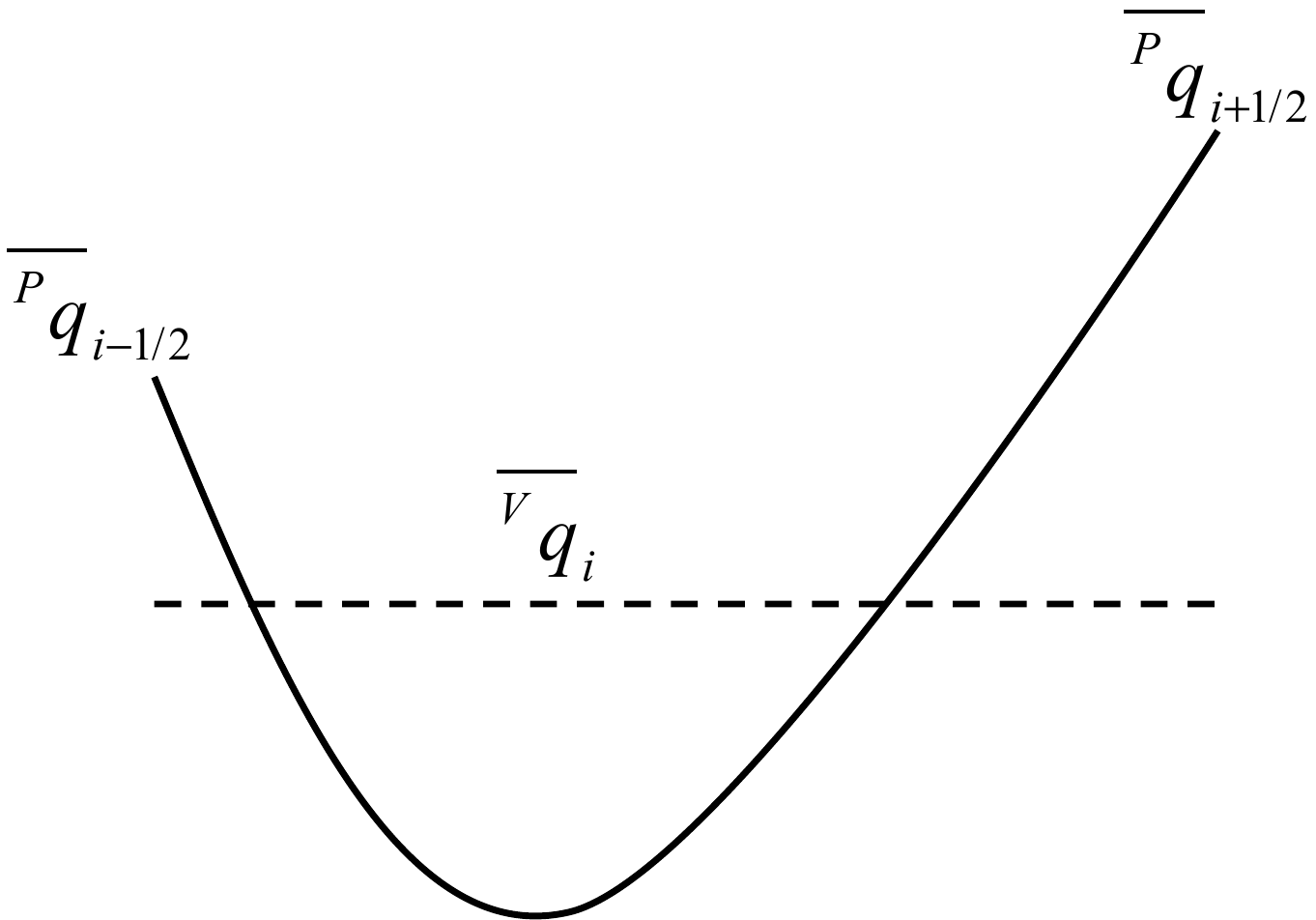}
\end{center}
\caption{Illustration of the rational reconstruction when a ‘valley’ is advected.}\label{rational figure}
\end{figure}

In addition, even the PV moments are all no less than 0, negative values may also appear when a ‘valley’ near lower boundary is advected. As illustrated in Fig. \ref{rational figure}, when the PV moments at cell boundary are bigger than the VIA moment, the reconstructed rational function would produce ‘undershoots’. In this condition, if the VIA is around lower boundary, negative values may appear. Similarly, when a ‘peak’ is advected, overshooting of PV moments may also appear. Thus, a further modification is needed

\begin{equation}
  R_i(x)=\overline{^V q}_{i}^{n+1} \quad \text { if } \overline{^V q}_{i}^{n+1} / q_{\max }<\varepsilon \text { and } \overline{^V q}_{i}^{n+1}<\overline{^P q}_{i\pm1 / 2}^{n+1} 
  \label{modification1}
\end{equation}

\begin{equation}
  R_i(x)=\overline{^V q}_{i}^{n+1} \quad \text { if } \overline{^V q}_{i}^{n+1} / q_{\max }>1-\varepsilon \text { and } \overline{^V q}_{i}^{n+1}>\overline{^P q}_{i\pm1 / 2}^{n+1}
  \label{modification2}
\end{equation}

where $\varepsilon$is a small parameter, such as $\varepsilon=10^{-3}$ , to avoid the ‘valley’ or ‘peak’ to be advected. We should note that, the modification of Eq.\eqref{modification1} can guarantee the spatial approximation profile is above 0, and by using the flux-form formula of VIA moment we can get an absolutely positive value. But around the upper boundary, by using Eq.\eqref{modification2} we can only get the spatial approximation profile below $q_{\max}$ , but cannot ensure the flux out is no less than the flux in, so overshooting of VIA moment around upper boundary may still exist. Thus by using these modification of PV moments, the numerical result can strictly preserve positive and reduce overshooting. 

In this paper, the scheme using Eq.\eqref{ri:new} as spatial reconstruction is called CSLR1, the scheme with two steps modification is called CSLR1-M. When  $\beta = 0$ in Eq.\eqref{ri:new}, the scheme becomes CSL2\cite{Yabe2001}.
Given the known point values and cell average values at previous time step, the updating procedure can be summarized as follows:

\begin{enumerate}[1)]
  \item Using Eq.\eqref{ri:new}, the spatial approximation of transport property on the whole domain can be determined.
  \item Point values are updated by Eq.\eqref{updatapv} and Eq.\eqref{cal:source term}.
  \item Cell averaged values are updated by Eq.\eqref{updatevia}.
  \item Modifying the PV moments by Eq.\eqref{modification0}, Eq.\eqref{modification1} and \eqref{modification2} to ensure a positive value and suppress overshoots at next time step.
\end{enumerate}

\section{Extending to the cubed-sphere grid}

In this section, we extend the CSLR1 and its modification CSLR1-M scheme to the cubed-sphere grid to develop a global transport model. The cubed-sphere grid we used in this paper is constructed by equiangular central projection, by this way six identical local coordinate $(\alpha, \beta)=[-\pi / 4, \pi / 4]$  are constructed.

The two-dimensional transport equation in local coordinate can be written as

\begin{equation}
  \frac{\partial(\sqrt{G} q)}{\partial t}+\frac{\partial\left(u^{1} \sqrt{G} q\right)}{\partial \alpha}+\frac{\partial\left(u^{2} \sqrt{G} q\right)}{\partial \beta}=0
\end{equation}

\noindent{where $\sqrt{G}$  is the Jacobian of transformation, and  $\left(u^{1}, u^{2}\right)$ is the contravariant components on the local coordinate, details can be found in\cite{Nair2005}. }

\begin{figure}[htbp]
\begin{center}
\includegraphics[width=0.6\textwidth]{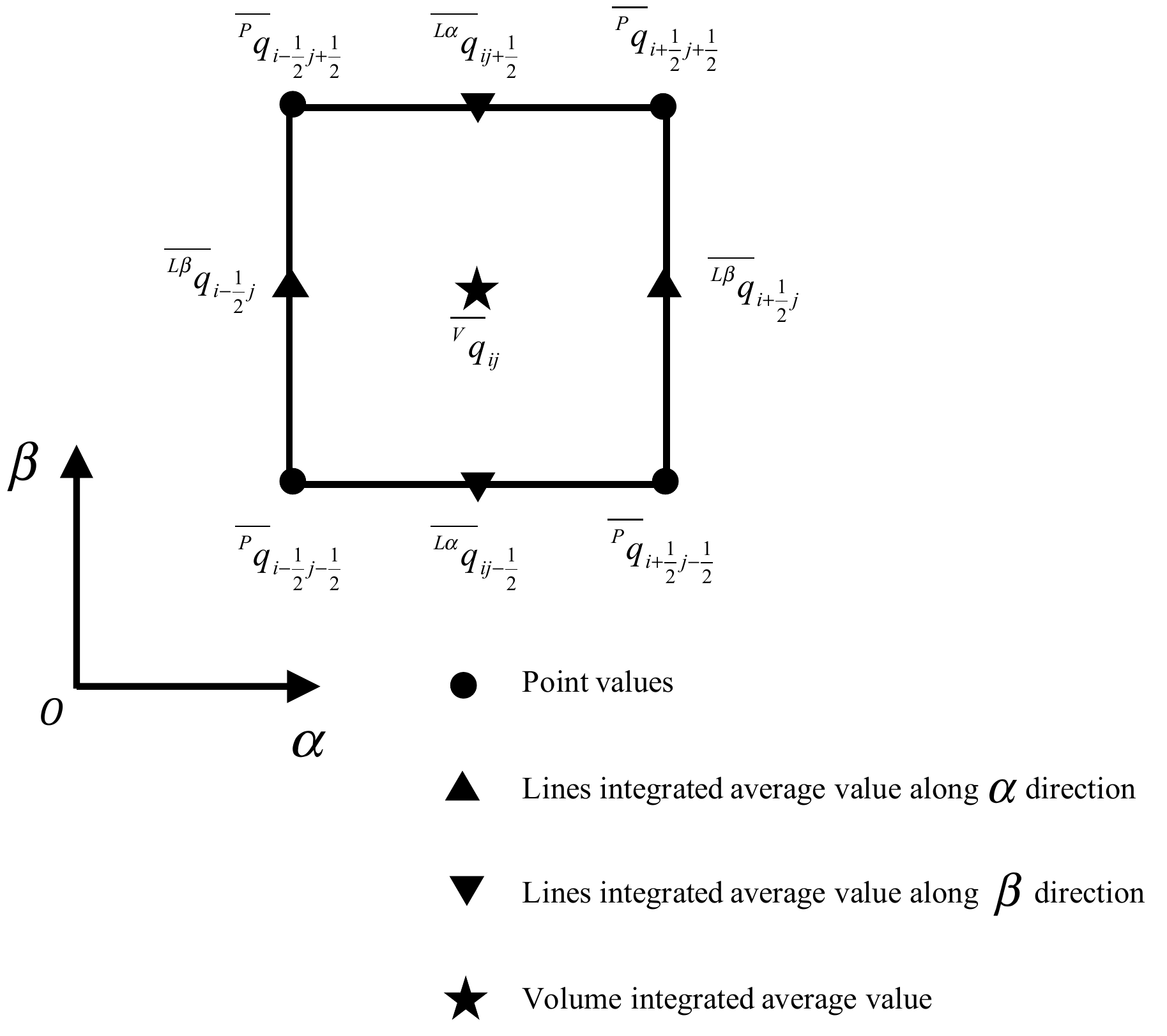}
\end{center}
\caption{Illustration of moments definition in two-dimensional case.}\label{illustraion2d}
\end{figure}

In the two-dimensional case, as shown in Fig.\ref{illustraion2d}, four kinds of moments are introduced within $C_{ij}$ :

\noindent{\textbullet \quad Volume integrated average (VIA):}

\begin{equation}
  \overline{^V q}_{i j}(t)=\frac{1}{\Delta \alpha \Delta \beta} \int_{\alpha_{i-1 / 2}}^{\alpha_{i+1/2}} \int_{\beta_{j-1/2}}^{\beta_{j+1/2}} q(\alpha, \beta, t) d \alpha d \beta
\end{equation}

\noindent{where $\Delta \alpha$  and $\Delta \beta$  are grid spacing in the $\alpha$  and $\beta$  direction respectively.}

\noindent{\textbullet \quad	Point value (PV): four point-values are located at the vertices}

\begin{equation}
  \overline{^P q}_{i \pm 1 / 2 j \pm 1 / 2}(t)=q\left(\alpha_{i \pm 1 / 2}, \beta_{j \pm 1 / 2}, t\right)
\end{equation}

\noindent{\textbullet \quad	Line integrated average values along $\alpha$  direction:}

\begin{equation}
  \overline{^{L \alpha} q}_{i j \pm 1 / 2}(t)=\frac{1}{\Delta \alpha} \int_{\alpha_{i-1/2}}^{\alpha_{i+1/2}} q\left(\alpha, \beta_{j \pm 1 / 2}, t\right) d \alpha
\end{equation}

\noindent{\textbullet \quad	Line integrated average values along $\beta$  direction:}

\begin{equation}
  \overline{^{L \beta} q}_{i \pm 1 / 2 j}(t)=\frac{1}{\Delta \beta} \int_{\beta_{j-1/2}}^{\beta_{j+1/2}} q\left(\alpha_{i \pm 1/2}, \beta, t\right) d \alpha
\end{equation}

\begin{figure}[htbp]
\begin{center}
\subfigure[]
{\includegraphics[width=0.43\textwidth]{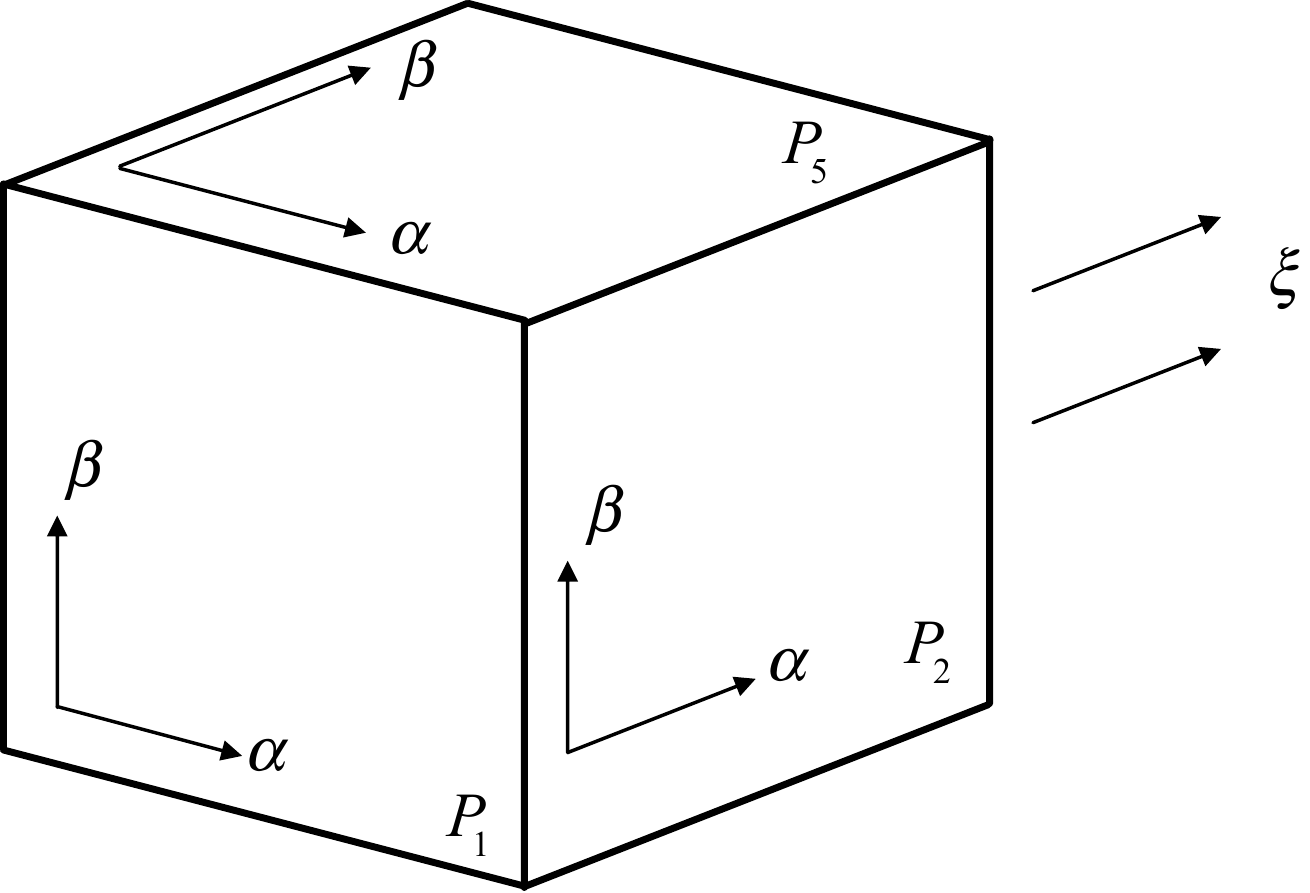}}
\subfigure[]
{\includegraphics[width=0.33\textwidth]{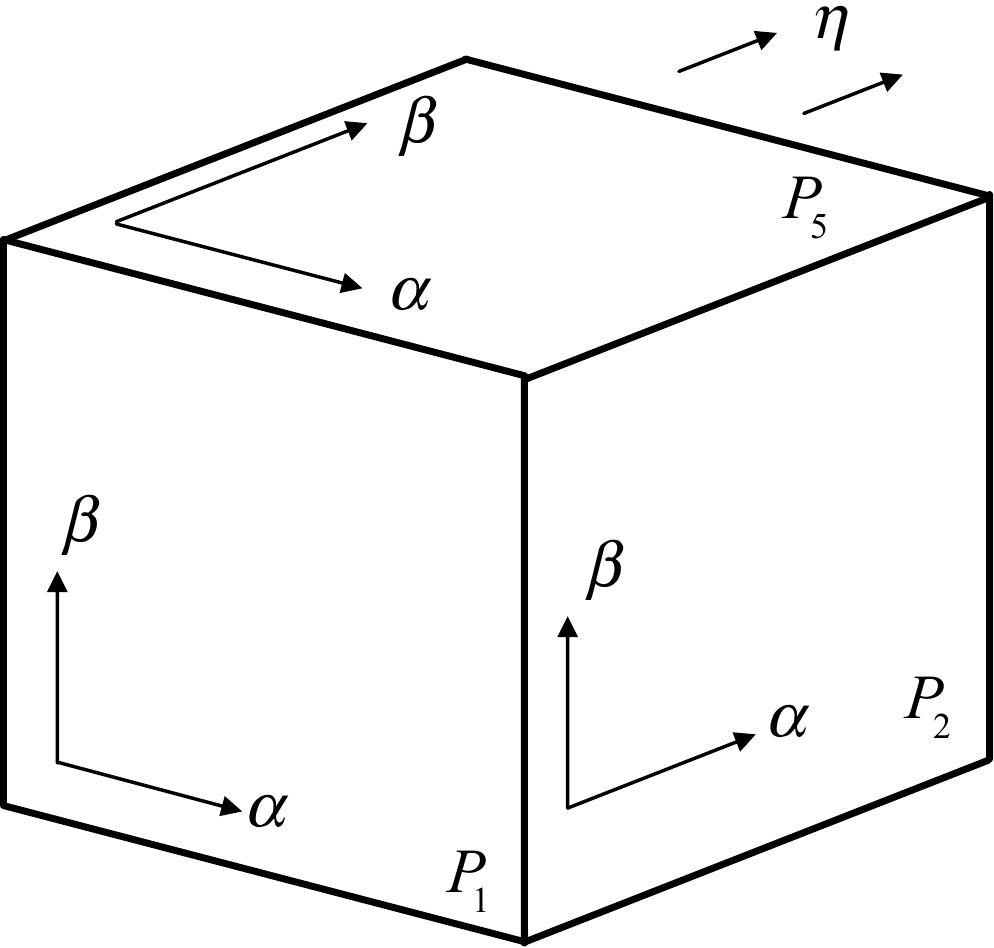}}
\subfigure[]
{\includegraphics[width=0.33\textwidth]{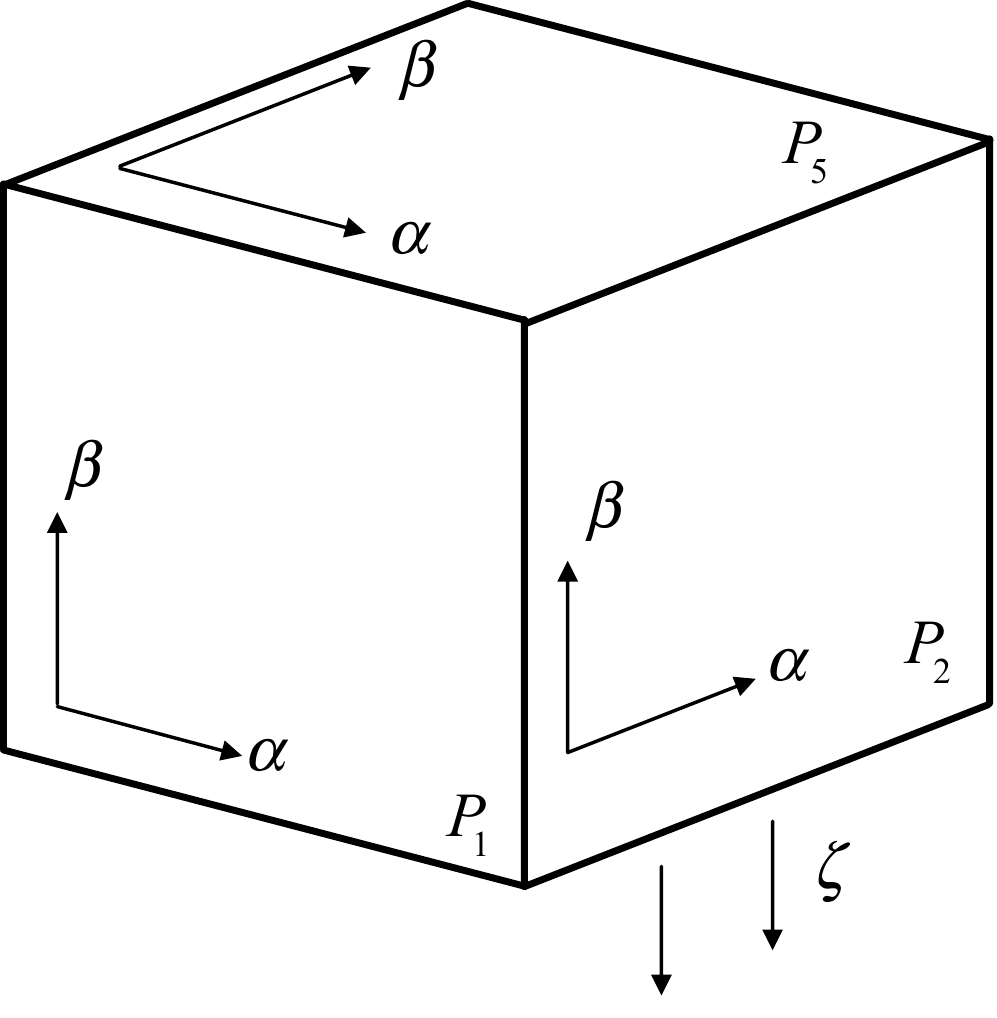}}
\end{center}
\caption{Schematic for three directions on the cubed-sphere grid.}\label{three direction}
\end{figure}

\begin{figure}[htbp]
\begin{center}
\includegraphics[width=0.4\textwidth]{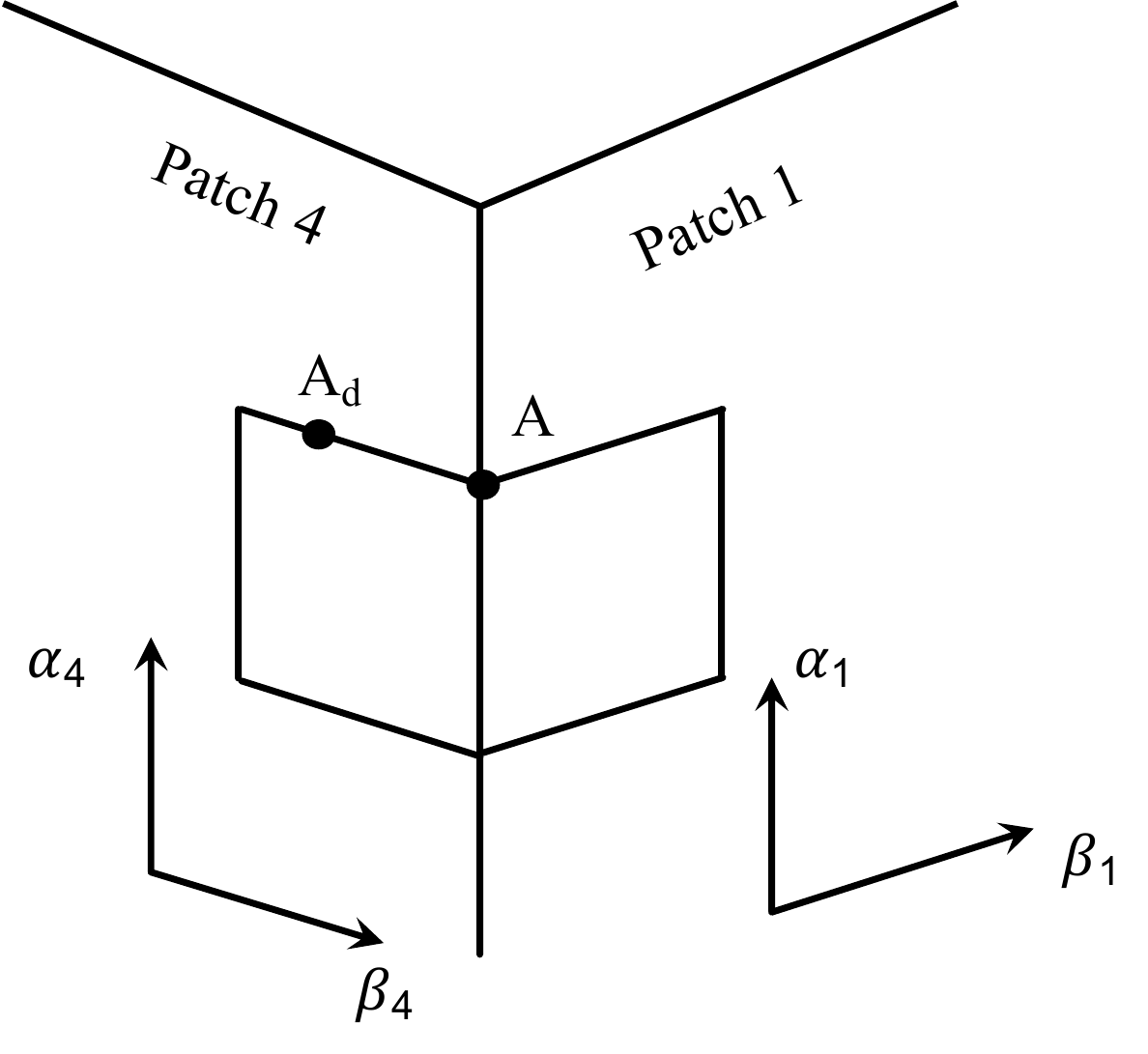}
\end{center}
\caption{Illustration of departure points along patch boundary. }\label{boundary}
\end{figure}

To ensure conservation property, we followed\cite{Guo2014} to divide the cube-sphere grid into three direction, see in Fig.\ref{three direction}. Firstly, we conduct the update procedure in $\xi$ direction, i.e. $\alpha$ direction on patch 1,2,3,4 for $\Delta t/2$. It should be noted that the moments along patch boundaries are only updated for once. As shown in Fig.\ref{boundary}, A is the arrival point on patch boundary and $\text{A}_{\text{d}}$ is the corresponding departure point on Patch 4, the point value of point A is calculated on Patch 4 and the flux across A is calculated by integrating the spatial approximation profile on Patch 4 along $\text{A}_{\text{d}}\text{A}$. If $\text{A}_{\text{d}}$ is on Patch 1, vice versa. By the same way, we update the moments in $\eta$  direction, i.e. $\beta$  direction on patch 1,3,5,6 for $\Delta t/2$ ; update in $\zeta$  direction, i.e. $\alpha$ direction on patch 5,6 and $\beta$  direction on patch 2,4 for $\Delta t$ ; then another $\Delta t/2$ in direction $\eta$ and $\xi$ direction separately to complete a full update procedure on the sphere geometry.

\section{Numerical experiments}

In this section, we use several widely used benchmark tests to verify the performance of the proposed transport model. Including solid body rotation, moving vortices and deformational flow test on the spherical mesh. 

The normalized errors and relative maximum and minimum errors proposed by Williamson et al.\cite{Williamson1992} are used:

\begin{equation}
  l_{1}=\frac{\int_{\Omega}\left|q-q_{t}\right| d \Omega}{\int_{\Omega}\left|q_{t}\right| d \Omega}
\end{equation}

\begin{equation}
  l_{2}=\sqrt{\frac{\int_{\Omega}\left(q-q_{t}\right)^{2} d \Omega}{\int_{\Omega} q_{t}^{2} d \Omega}}
\end{equation}

\begin{equation}
  l_{\infty}=\frac{\max \left|q-q_{t}\right|}{\max \left|q_{t}\right|}
\end{equation}

\begin{equation}
  q_{\max }=\frac{\max (q)-\max \left(q_{t}\right)}{\max \left(q_{t}\right)-\min \left(q_{t}\right)}
\end{equation}

\begin{equation}
  q_{\min }=\frac{\min (q)-\min \left(q_{t}\right)}{\max \left(q_{t}\right)-\min \left(q_{t}\right)}
\end{equation}

\noindent{where $\Omega$ is the whole computational domain, $q$ and $q_t$ refer to numerical solutions (volume integrated average in our paper) and exact solutions respectively.}

\subsection{ Solid-body rotation tests}

Solid-body rotation test\cite{Williamson1992} is widely used in two-dimensional spherical transport model to evaluate the performance of a transport model. The wind components in the latitude-longitude coordinates $(\lambda,\theta)$ are defined as 

\begin{equation}
  u_{s}(\lambda, \theta)=u_{0}(\cos \theta \cos \alpha+\sin \theta \cos \lambda \sin \alpha)\label{solid_u}
\end{equation}

\begin{equation}
  v_{s}(\lambda, \theta)=-u_{0} \sin \lambda \sin \alpha
  \label{solid_v}
\end{equation}

\noindent{where $\left( u_s,v_s \right)$ is the velocity vector, $u_0=2\pi R/12\text{days}$, which means it takes 12 days to complicate a full rotate on the sphere, $R$ is the radius of the sphere, and $\alpha$ is a parameter which controls the rotation angle. In this test, two kinds of initial conditions are used, including a cosine bell and a step cylinder.}

(a) Solid body rotation of a cosine bell

The initial condition of a cosine bell test is specified as

\begin{equation}
  q(\lambda, \theta, 0)=\left\{\begin{array}{ll}
    \left(h_{0} / 2\right)\left[1+\cos \left(\pi r_{d} / r_{0}\right)\right] & \text { if } r_{d}<r_{0} \\
    0 & \text { if } r_{d} \geq r_{0}
    \end{array}\right.
\end{equation}

\noindent{where $r_d$ is the great circle distance between $(\lambda,\theta)$  and the center of the cosine bell, located at $(3\pi /2,0)$, $r_0=7\pi R/64$ is the radius of the cosine bell, $h_0=1$.}

The normalized errors on $30\times 30\times 6$ meshes and with 256 time-steps compared with other existing published Semi-Lagrangian scheme, the PPM-M scheme\cite{Zerroukat2007} and CSLAM-M\cite{Lauritzen2010}, are presented in Table \ref{test1.1}. With a qusi-smooth initial condition and quiet simple velocity field, the modification procedure has little effect on the numerical result. The result shows that CSLR1 and CSLR1-M has almost the same result. And our scheme is comparable to PPM-M scheme, the result in near pole flow direction ($\alpha=\pi /2$ and $\alpha=\pi /2-0.05$) is better than CSLAM-M scheme. 

\begin{figure}[htbp]
\begin{center}
\includegraphics[width=0.6\textwidth]{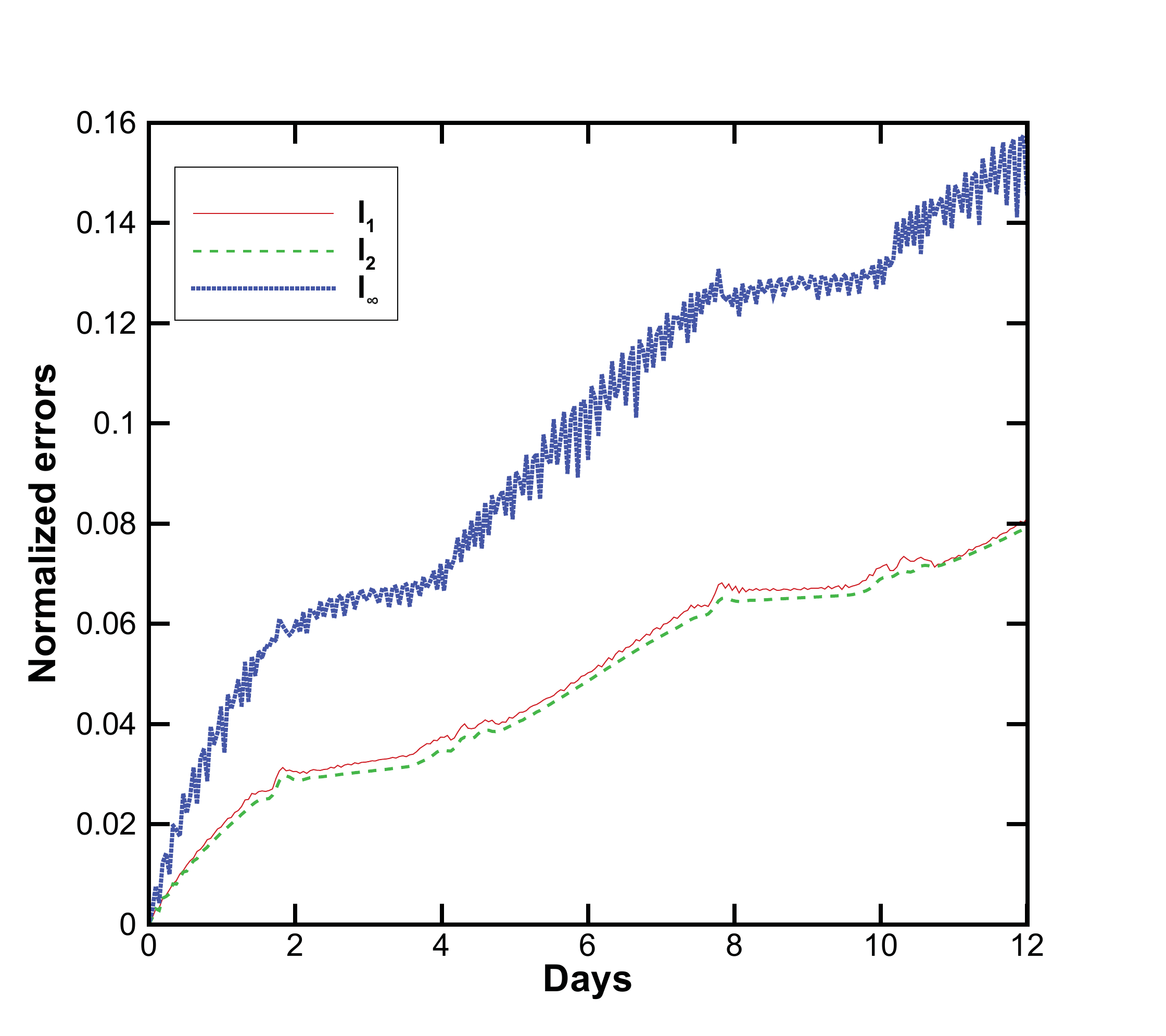}
\end{center}
\caption{History of normalized errors of the solid body rotation of a cosine bell for one revolution on grid $N=32$(number of cells in one direction on each cell), 256 time-steps and with $\alpha=\pi /4$.}\label{normalized error history}
\end{figure}

To check the influence of the weak singularities at 8 the vertices of the cubed-sphere gird, this test is conducted with $\alpha=\pi /4$ to pass through four vertices. The history of normalized errors (CSLR1 and CSLR1-M are almost the same, here we only present the result of CSLR1-M) are shown in Fig.\ref{normalized error history}. We can see that the normalized errors have no visible fluctuations when the flow passes four weak singularities. 

\begin{table}[htbp]
\caption{Comparison the normalized errors of rotation of a cosine bell after one revolution with other published schemes.} \label{test1.1}
\begin{tabularx}{\textwidth}{l@{\extracolsep{\fill}}ccc}
\hline
\multirow{2}{*}{Scheme}
&\multicolumn{3}{c}{$\alpha=0$}\\
\cline{2-4}
&{$l_1$}  &{$l_2$}  &{$l_\infty$} \\
\hline
CSLR1(CSLR1-M) & 0.116 &	0.097 &	0.112\\
PPM-M	& 0.101 &	0.095 &	0.115\\
CSLAM-M &	0.075 &	0.075 &	0.141\\
\hline
\multirow{2}{*}{Scheme}
&\multicolumn{3}{c}{$\alpha=\pi /4$}\\
\cline{2-4}
&{$l_1$}  &{$l_2$}  &{$l_\infty$} \\
\hline
CSLR1(CSLR1-M) &	0.081	& 0.079 &	0.145\\
PPM-M &	0.078 &	0.086	& 0.159\\
CSLAM-M &	0.048 &	0.060 &	0.130\\
\hline
\multirow{2}{*}{Scheme}
&\multicolumn{3}{c}{$\alpha=\pi /2$}\\
\cline{2-4}
&{$l_1$}  &{$l_2$}  &{$l_\infty$} \\
\hline
CSLR1(CSLR1-M) &	0.082	& 0.072	& 0.084\\
PPM-M	& 0.109 &	0.102 &	0.118\\
CSLAM-M &	0.075 &	0.075 &	0.141\\
\hline
\multirow{2}{*}{Scheme}
&\multicolumn{3}{c}{$\alpha=\pi /2-0.05$}\\
\cline{2-4}
&{$l_1$}  &{$l_2$}  &{$l_\infty$} \\
\hline
CSLR1(CSLR1-M) &	0.082 &	0.072 &	0.094\\
PPM-M	& 0.109 &	0.102 &	0.124\\
CSLAM-M &	0.070 &	0.069 &	0.133\\
\hline
\end{tabularx}
\end{table}

To demonstrate the ability of using large CFL number to transport, we use 72 time-steps to complete one revolution. The normalized errors compared with CSLAM-M are shown in Table \ref{test1.2}, the proposed scheme in this paper has larger $l_1$ and $l_2$ error, while $l_{\infty}$ error is smaller than CSLAM-M. Compare with the result using 256 time-steps with the same grid resolution and flow direction, a larger CFL number condition even get a better result in terms of norm errors. 

\begin{table}[htbp]
  \caption{. Same as Table 1, but with 72 time-steps} \label{test1.2}
  \begin{tabularx}{\textwidth}{l@{\extracolsep{\fill}}ccc}
  \hline
  \multirow{2}{*}{Scheme}
  &\multicolumn{3}{c}{$\alpha=\pi /2$}\\
  \cline{2-4}
  {$l_1$}  &{$l_2$}  &{$l_\infty$} \\
  CSLR1	& 0.062 &	0.049 &	0.058\\
  CSLR1-M	& 0.062 &	0.050 &	0.064\\
  CSLAM-M &	0.029 &	0.033 &	0.070\\
  \hline
  \end{tabularx}
  \end{table}

  (b) Solid body rotation of a step cylinder

  A non-smooth step cylinder is calculated to evaluate the non-oscillatory property. The initial distribution is specified as

  \begin{equation}
    q(\lambda, \theta, 0)=\left\{\begin{array}{ll}
      1000 & \text { if } r_{d}<r_{1} \\
      500 & \text { if } r_{1} \leq r_{d}<r_{2} \\
      0 & \text { if } r_{d} \geq r_{2}
      \end{array}\right.
  \end{equation}

  \noindent{where $r_d$ is the great circle distance between $(\lambda,\theta)$ and $(3\pi /2,0)$, which is the center of the step cylinder, $r_1=2/3R$ and $r_2=1/3R$.}

\begin{figure}[htbp]
\begin{center}
\subfigure[The result of CSL2]
{\includegraphics[width=0.3\textwidth]{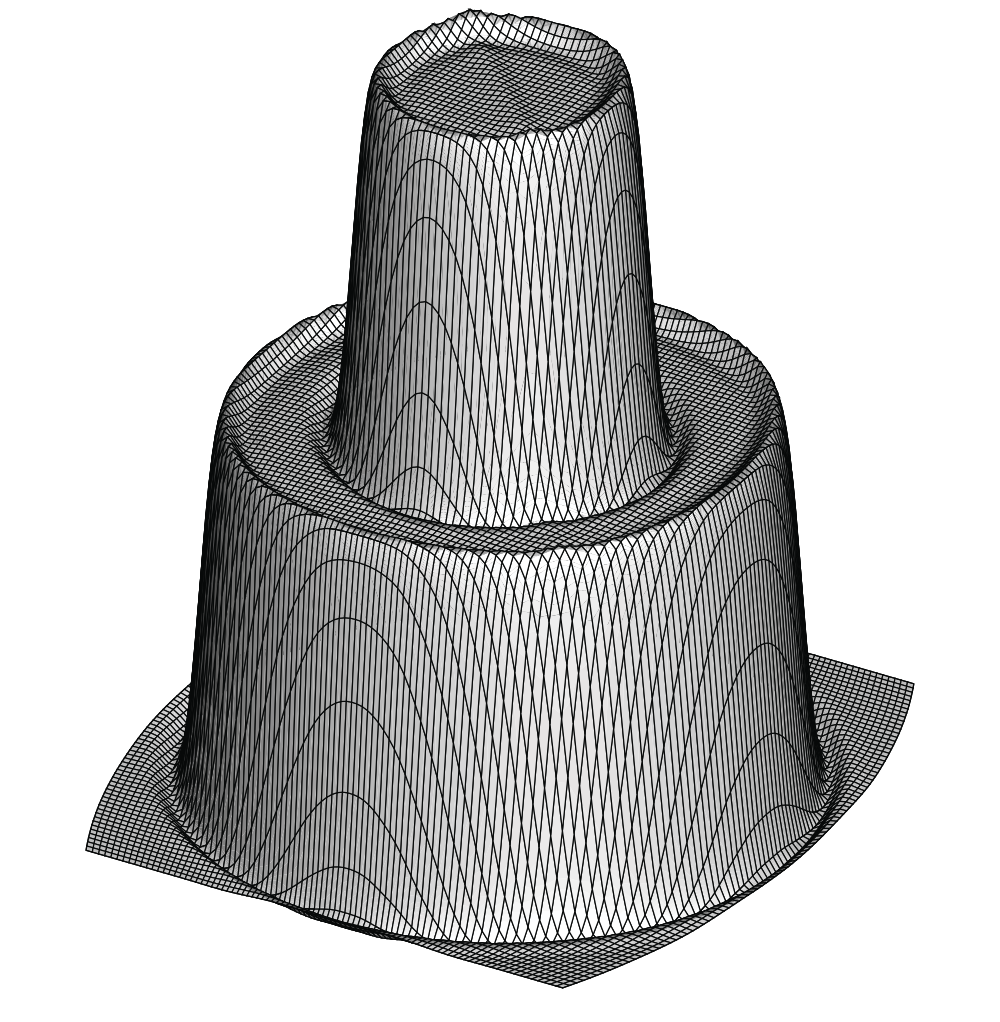}}
\subfigure[The result of CSLR1]
{\includegraphics[width=0.3\textwidth]{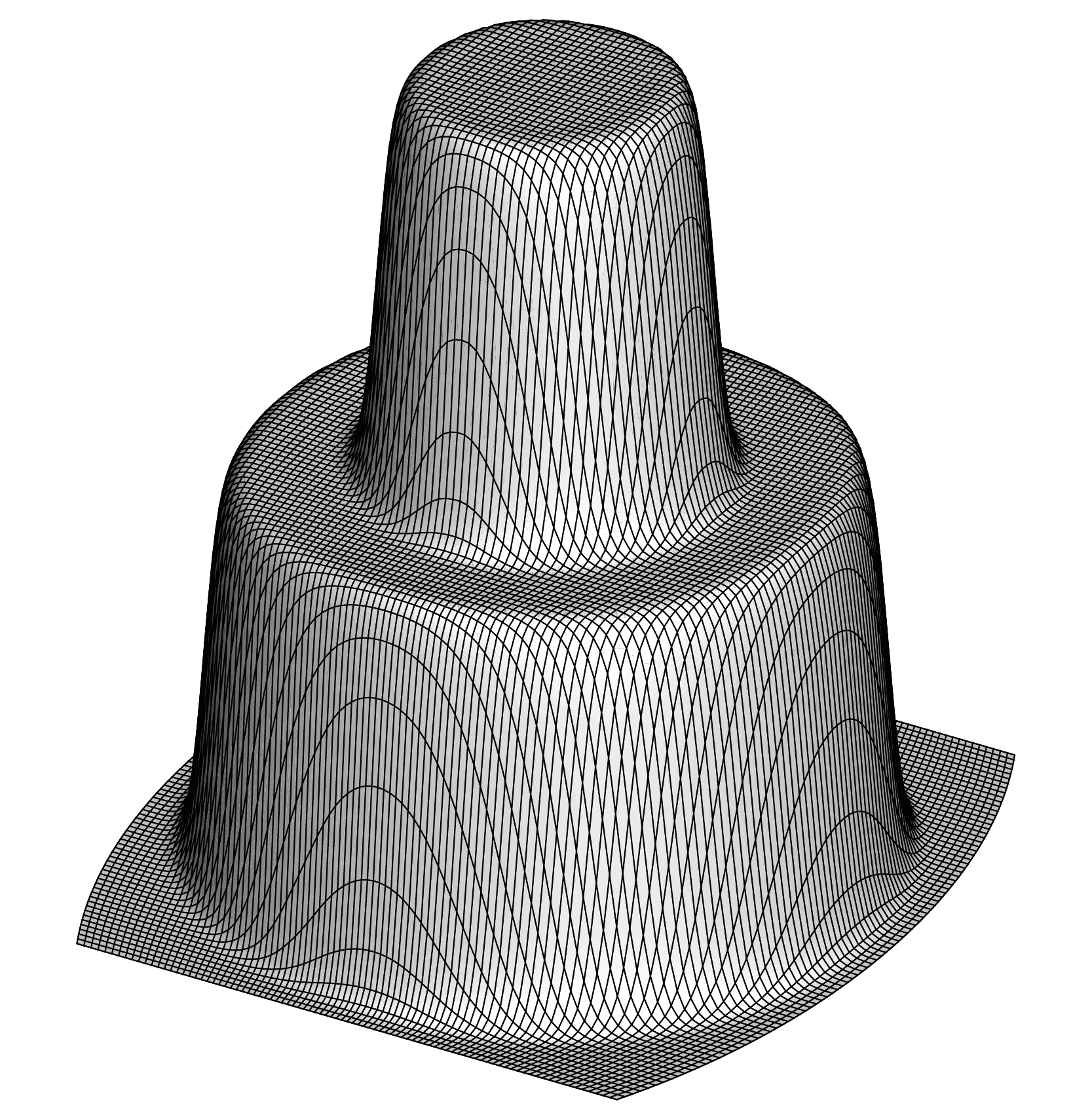}}
\subfigure[The result of CSLR1-M]
{\includegraphics[width=0.3\textwidth]{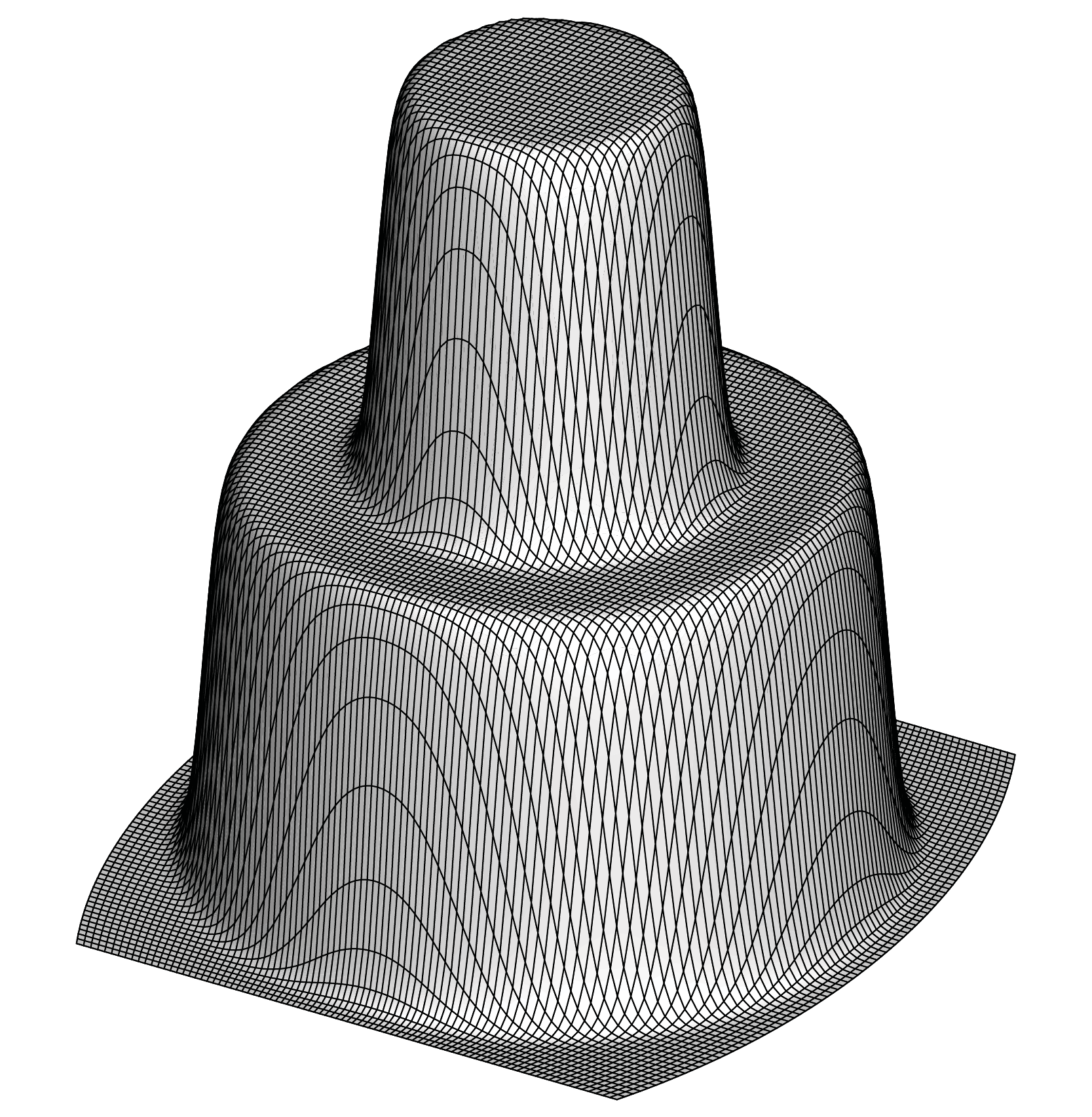}}
\end{center}
\caption{Numerical result of solid body rotation of step cylinder after one revolution (12 days). }\label{stepcylinder}
\end{figure}

\begin{figure}[htbp]
\begin{center}
\includegraphics[width=0.6\textwidth]{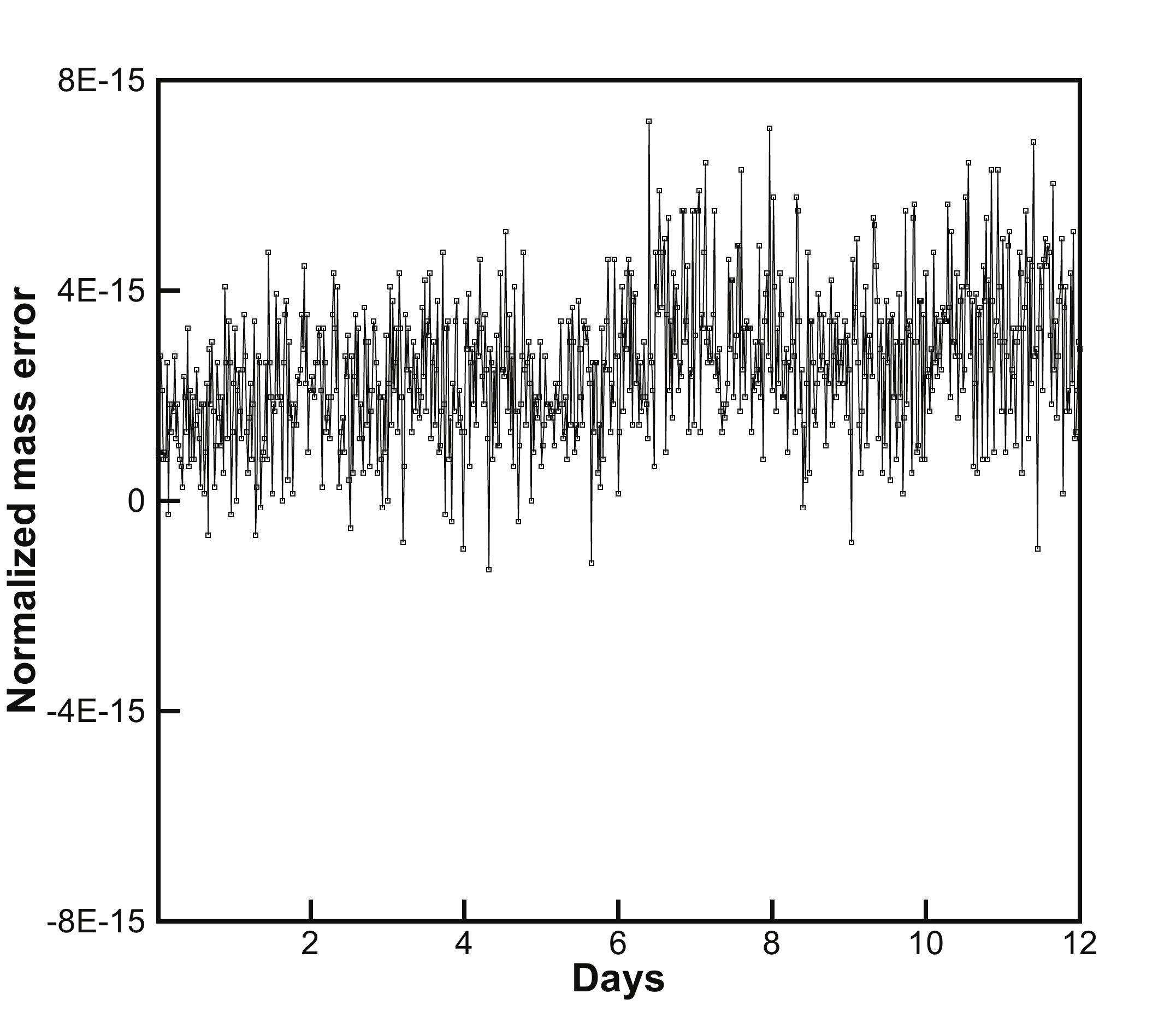}
\end{center}
\caption{The time history of total normalized mass error for solid body rotation test of CSLR1-M scheme.}\label{masserror}
\end{figure}

  In this test, we set $\alpha=\pi /4$, which is the most challenging case, the step cylinder moves through four vertices and two edges to complete a full rotation. Here we use $90 \times 90 \times 6$ meshes and with 720 steps to conduct this test. The numerical results after 12 days are shown in Fig. \ref{stepcylinder}, we can see that the CSL2 scheme will generate oscillations around the discontinuities. But by using the CLSR1 and CSLR1-M approach, these unphysical oscillations are effectively removed. The relative maximum and minimum of CSL2 are $q_{\max} =3.77 \times 10^{-2}$ and $q_{\min}=-2.80 \times 10^{-2}$, for CLSR1 it is  $q_{\max} =1.36 \times 10^{-3}$ and $q_{\min}=0$, for CSLR1-M is $q_{\max} =-9.23 \times 10^{-5}$ and $q_{\min}=0$. The result indicates that CSLR1-M can effectively remove overshoots in this test. The history of normalized mass errors is given in Fig. \ref{masserror}, which shows that the normalized mass error is within the tolerance of machine precision, therefore the proposed global transport model is exactly mass conservative during simulation procedure.

\subsection{ Moving vortices on the sphere}

The second benchmark test we used is the moving vortices proposed by Nair et al.\cite{Nair2008}, the wind component of which is a combination of rotation test and two vortices, it is much more complicated than the solid-body rotation test. The velocity field on the sphere is specified as

\begin{equation}
  u(\lambda, \theta, t)=u_{s}(\lambda, \theta)+R \omega_{r}\left[\sin \theta_{c}(t) \cos \theta-\cos \theta_{c}(t) \cos \left(\lambda-\lambda_{c}(t)\right) \sin \theta\right]
\end{equation}

\begin{equation}
  v(\lambda, \theta, t)=v_{s}(\lambda, \theta)+R \omega_{r}\left[\cos \theta_{c}(t) \sin \left(\lambda-\lambda_{c}(t)\right)\right]
\end{equation}

\begin{equation}
  \omega_{r}\left(\theta_{c}(t)\right)=\frac{V}{r \rho}
\end{equation}

\begin{equation}
  V(\rho)=u_{0} \frac{3 \sqrt{3}}{2} \operatorname{sech}^{2} \rho \tanh \rho
\end{equation}

\begin{equation}
  \rho\left(\theta_{c}(t)\right)=\rho_{0} \cos \theta_{c}(t)
\end{equation}

where $u_s$ and $v_s$ are calculated by Eq.\eqref{solid_u} and \eqref{solid_v}, the rotation angle of this test is set to be $\alpha=\pi /4$. $\rho _0=3$, $\lambda_c(t)$ and  $\theta_c(t)$ are the center of the moving vortex at time t, the calculation procedure of $\lambda_c(t)$ and  $\theta_c(t)$ can be found in \cite{Nair2008}.

The tracer field is defined as 

\begin{equation}
  q\left(\lambda^{\prime}, \theta^{\prime}, t\right)=1-\tanh \left(\frac{\rho}{\gamma} \sin \left(\lambda^{\prime}-\omega_{r} t\right)\right)\label{ee}
\end{equation}

\noindent{where $\gamma$ is a parameter to control the smoothness of tracer field, $(\lambda \prime,\theta \prime)$ is the rotated spherical coordinates, which can be calculated by}

\begin{equation}
  \lambda^{\prime}(\lambda, \theta)=\tan ^{-1}\left(\frac{\cos \theta \sin \left(\lambda-\lambda_{p}\right)}{\cos \theta \sin \theta_{p} \cos \left(\lambda-\lambda_{p}\right)-\cos \theta_{p} \sin \theta}\right)
\end{equation}

\begin{equation}
  \theta^{\prime}(\lambda, \theta)=\sin ^{-1}\left(\sin \theta \sin \theta_{p}+\cos \theta \cos \theta_{p} \cos \left(\lambda-\lambda_{p}\right)\right)
\end{equation}

\noindent and $\left(\lambda_{p}, \theta_{p}\right)=(\pi, \pi / 2-\alpha)$ is the North Pole of the rotated spherical coordinate. In this test, we followed\cite{Norman2018} to set $\gamma =10^{-2}$ to conduct a large gradient in tracer distribution. When $t=0$ in Eq. \eqref{ee}, we get the initial condition. 

\begin{figure}[htbp]
\begin{center}
\subfigure[Exact solution]
{\includegraphics[width=0.6\textwidth]{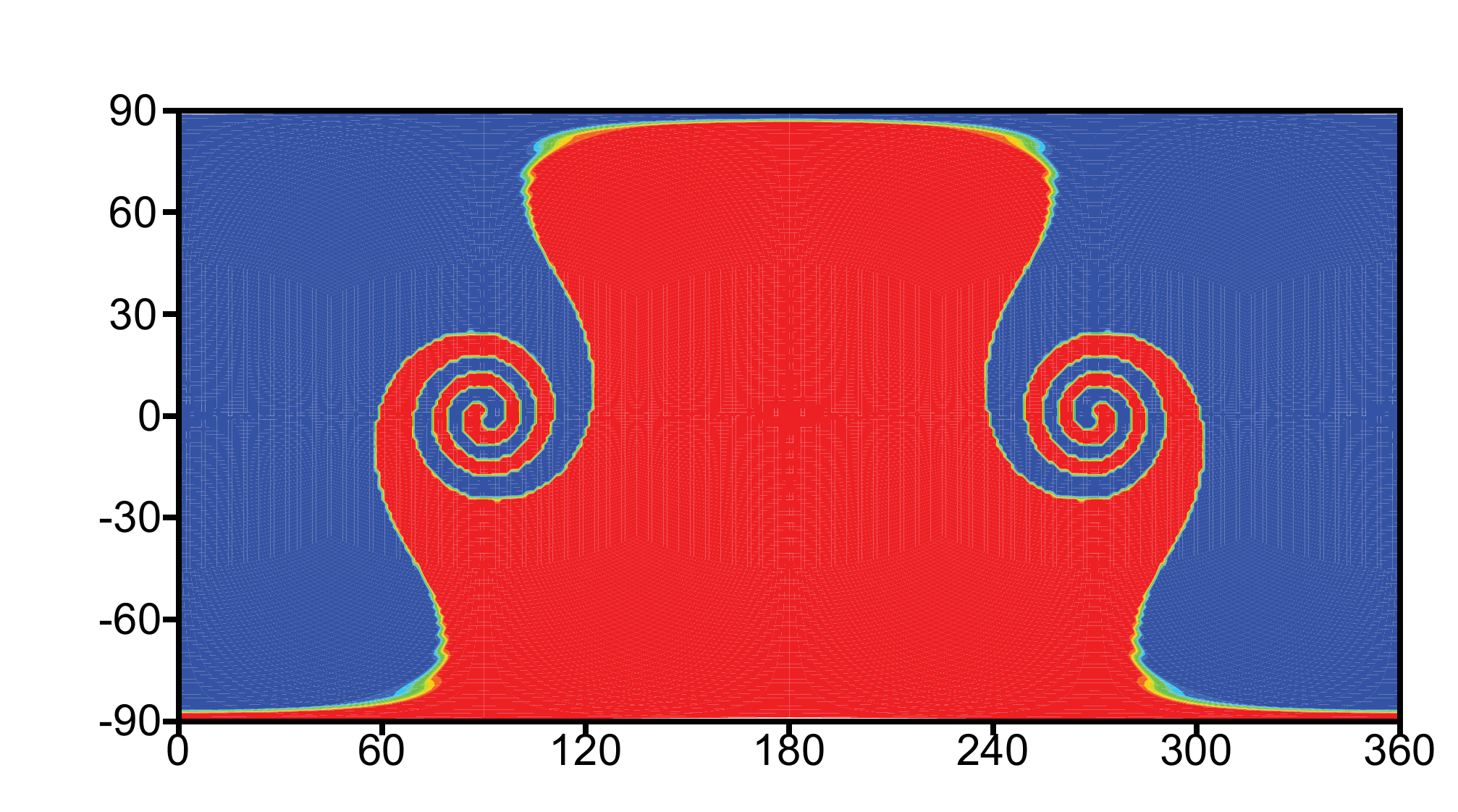}}
\subfigure[Numerical result of CSLR1-M]
{\includegraphics[width=0.6\textwidth]{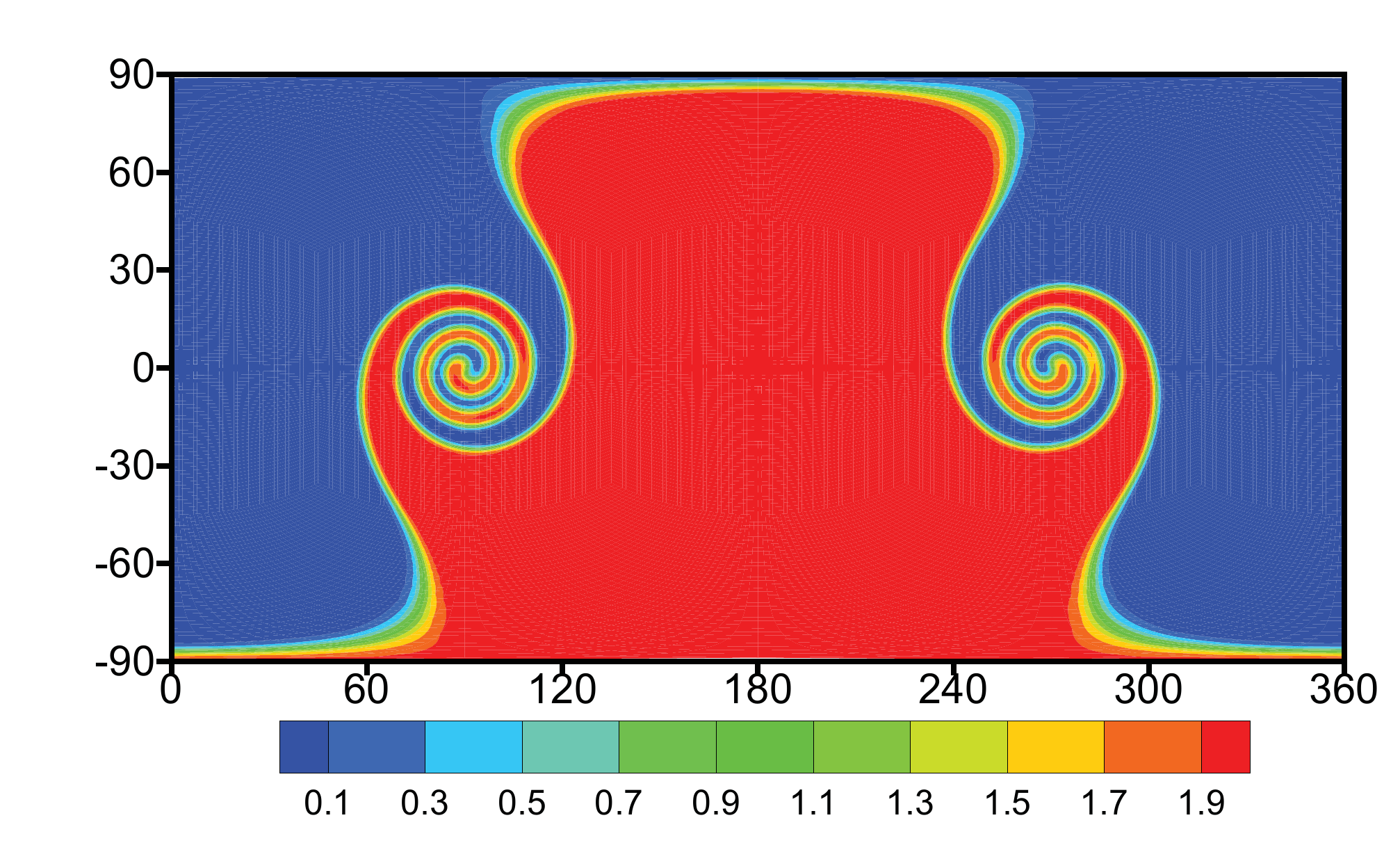}}
\end{center}
\caption{Contour plot of moving vortices after 12 days.}\label{moving}
\end{figure}

The test is conducted on $80\times 80 \times 6$ grid and uses 400 steps to move for 12 days. Contour plot is shown in Fig. \ref{moving}, compared with the exact solution, the proposed scheme can correctly simulate this complicated procedure. The errors of this test are presented in Table \ref{errormoving}, we can see that the normalized errors $l_1$,$l_2$, $l_{\infty}$ and $q_{\max}$ of CSLR1 and CSLR1-M have little difference, CSLR1 scheme have small undershoots, while CSLR1-M can completely preserve non-negativity. 

\begin{table}[htbp]
\caption{Errors of moving vortices.} \label{errormoving}
\begin{tabularx}{\textwidth}{l@{\extracolsep{\fill}}ccccc}
\hline\hline
Scheme & $l_1$ & $l_2$ & $l_{\infty}$ & $q_{\max}$& $q_{\min}$\\ 
\hline
CSLR1	& 0.0559 &	0.1402 &	0.8170 &	1.2560e-2	& -3.3841e-6\\
CSLR1-M	& 0.0569 &	0.1402 &	0.8162 &	1.7448e-2 &	0.0000\\
\hline
\end{tabularx}
\end{table}

\subsection{Deformational flow test}

The last benchmark test used in our paper is deformational flow test proposed by Nair and Lauritzen\cite{Nair2010}, which is a very challenging test case. The flow field is nondivergent and time-dependent:

\begin{equation}
  u(\lambda, \theta, t)=\kappa \sin ^{2} \lambda^{\prime} \sin (2 \theta) \cos \left(\frac{\pi t}{T}\right)+\frac{2 \pi}{T} \cos \theta
\end{equation}

\begin{equation}
  v(\lambda, \theta, t)=\kappa \sin \lambda^{\prime}\cos \theta \cos \left( \frac{\pi t}{T} \right)
\end{equation}

\noindent where $\kappa =2$, $T=5$, and $\lambda^{\prime}=\lambda-(2 \pi t/T)$.

Two kinds of initial conditions are checked here, including twin slotted cylinder case to evaluate the positivity preserving property and correlated cosine bells to evaluate the nonlinear correlations between tracers\cite{Lauritzen2012}. By the given flow field, the initial distributions will change into thin bars in the first half period, then return to its initial state in the second half period.

(a) Deformation of twin slotted cylinder

The initial condition is defined as

\begin{equation}
  q(\lambda, \theta, 0)=\left\{\begin{array}{ll}
    1 & \text { if } \ r_i\leq r_0;|\lambda - \lambda_i| \geq \frac{r_0}{6};i=1,2\\
    1 & \text { if } \ r_1 \leq r_0;|\lambda - \lambda_1| < \frac{r_0}{6}; \theta-\theta_1 < -\frac{5r_0}{12} \\
    1 & \text { if } \ r_2 \leq r_0;|\lambda - \lambda_2| < \frac{r_0}{6}; \theta-\theta_1 > -\frac{5r_0}{12} \\
    0 & \text{otherwise}
    \end{array}\right.
\end{equation}

\noindent where $r_0=0.5$ and $r_i(i=1,2)$ means the great circle distance between the center of the two slotted cylinder and a given point. The center of the two slotted cylinders are located at $(\lambda_1,\theta_1)=(5 \pi /6,0)$ and $(\lambda_2,\theta_2)=(7 \pi /6,0)$, respectively.

\begin{figure}[htbp]
\begin{center}
\subfigure[]
{\includegraphics[width=0.6\textwidth]{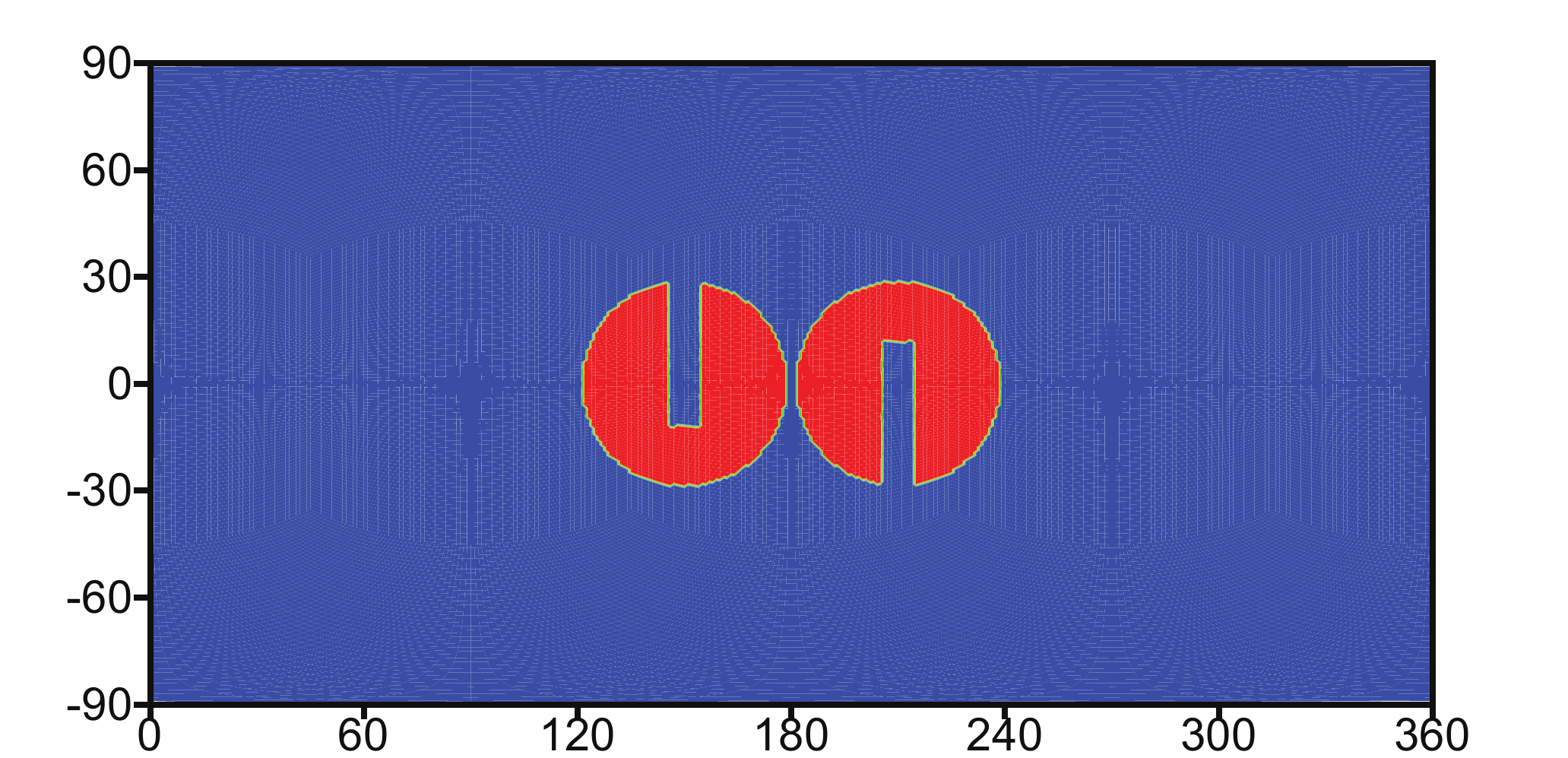}}
\subfigure[]
{\includegraphics[width=0.6\textwidth]{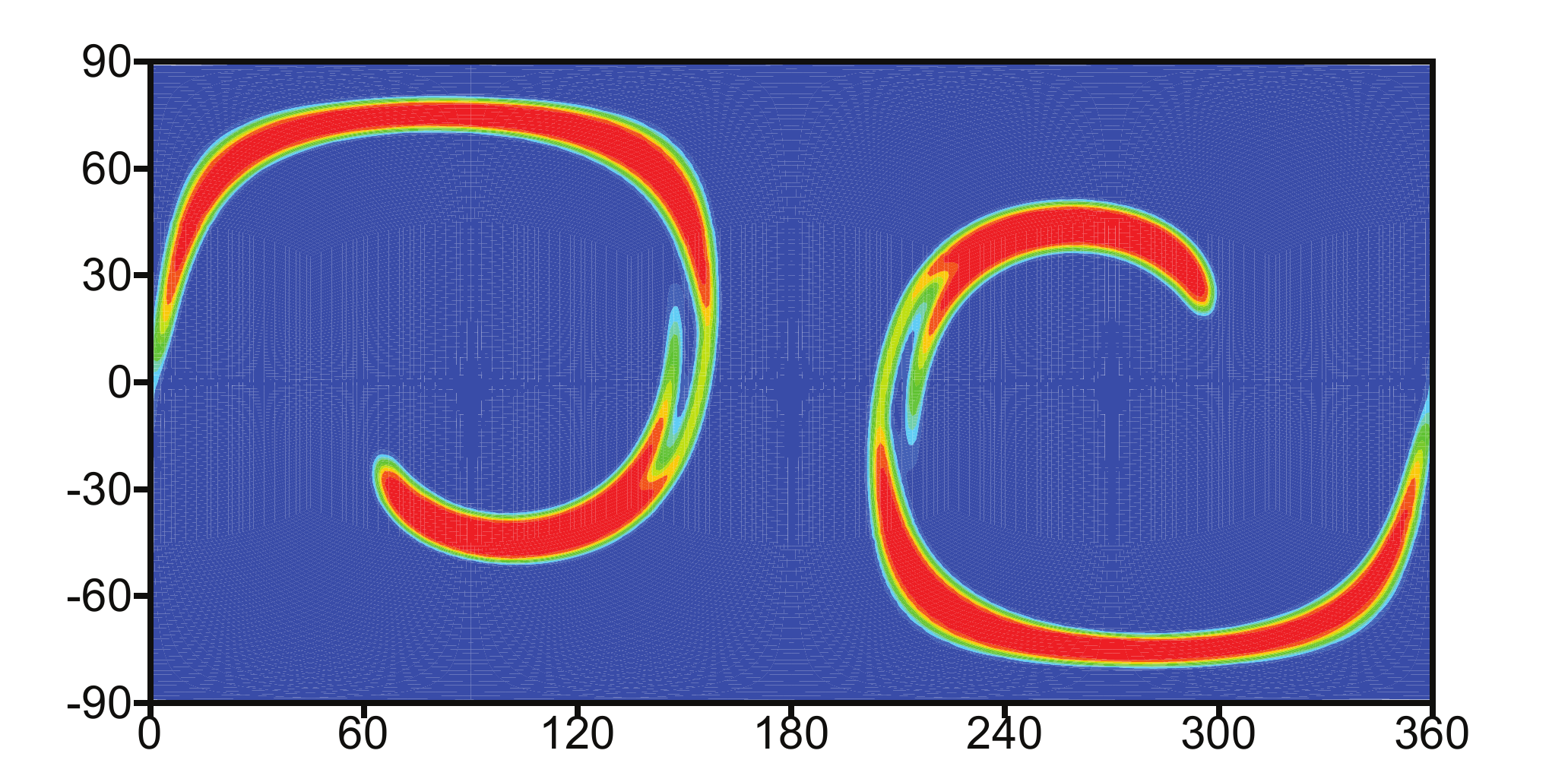}}
\subfigure[]
{\includegraphics[width=0.6\textwidth]{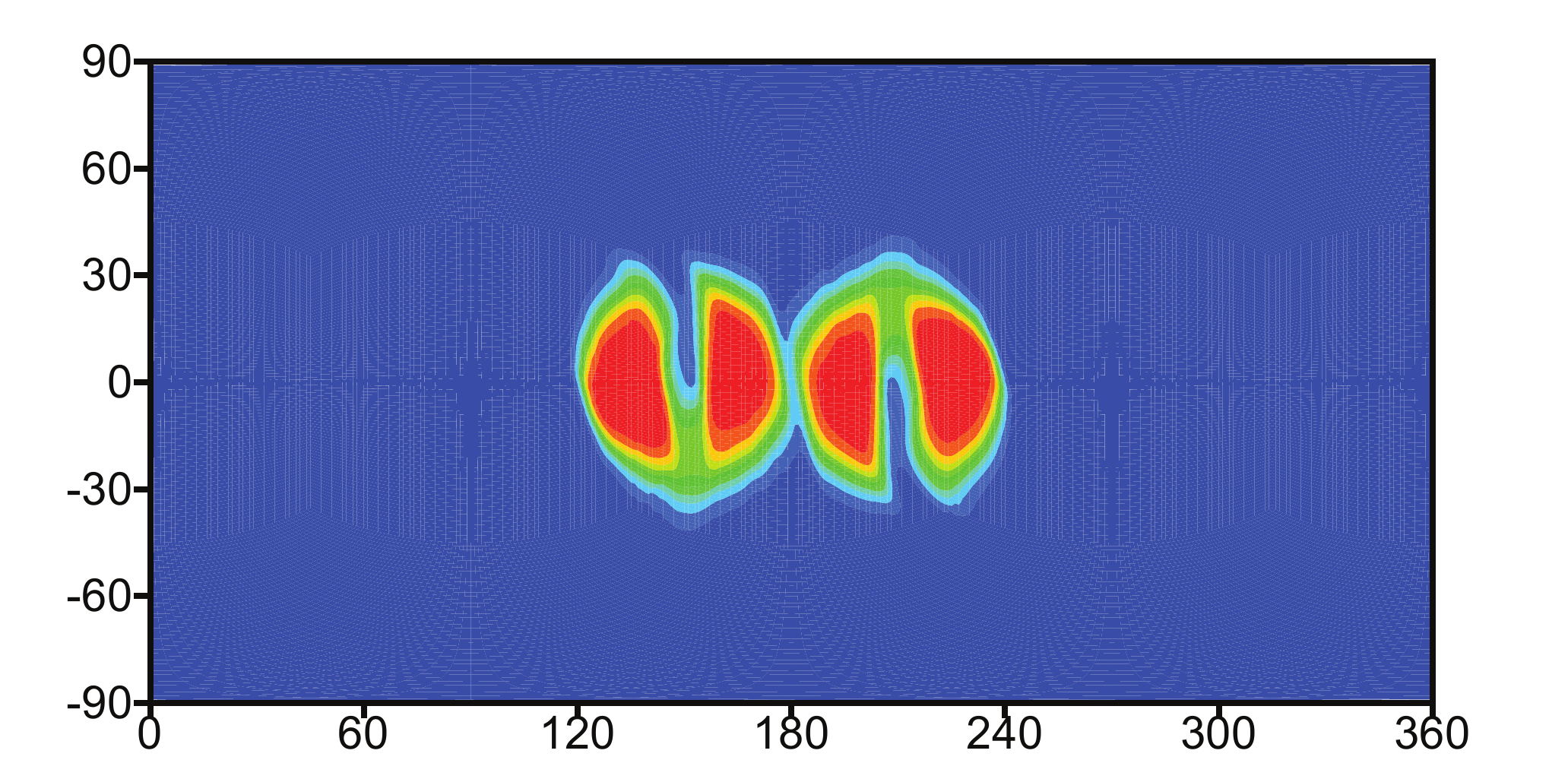}}
\end{center}
\caption{Numerical result of deformational flow of slotted cylinder after one period. (a) is the exact solution, (b) the numerical solution at half cycle, (c) the numerical solution after one cycle.}\label{deformation}
\end{figure}

The numerical result of deformational flow of CSLR1-M scheme on $90 \times 90 \times 6$ grid and with 390 time-steps (local $\text{CFL}_{\max}$ is about 3) is shown in Fig. \ref{deformation}. The result at first half period is shown in Fig. \ref{deformation}(b), the two slotted cylinders are deformed into two thin filaments by the background flow field. Fig. \ref{deformation}(c) is result at final time, it is clearly that the proposed scheme can correctly reproduce this complicate deformational flow. Normalized errors are shown in Table \ref{errorslotted}. Both schemes are non-negative at final step, but the CSLR1 scheme will generate negative values at some time step, while CSLR1-M can always preserve positive along the simulation procedure. And the CSLR1-M scheme reduces the overshooting slightly.

\begin{table}[htbp]
  \caption{Errors of deformational flow of twin slotted cylinder case.} \label{errorslotted}
  \begin{tabularx}{\textwidth}{l@{\extracolsep{\fill}}ccccc}
  \hline
  Scheme & $l_1$ & $l_2$ & $l_{\infty}$ & $q_{\max}$& $q_{\min}$\\ 
  \hline
  CSLR1 &	0.3080 &	0.3118 &	0.7142 &	5.4986e-2 &	0.0000\\
  CSLR1-M &	0.3092 &	0.3116 &	0.7142 &	8.2343e-3 &	0.0000\\
  \hline
  \end{tabularx}
\end{table}

  (b) Deformation of cosine bell test

  We replace the twin slotted cylinder with the quasi-smooth twin cosine bells:

  \begin{equation}
    q^{\text{cb}}=\left\{\begin{array}{ll}
      0.1+0.9h_i(\lambda,\theta) & \text {if} \ r_i<r_0(i=1,2)\\
      0.1 & \text{otherwise}
      \end{array}\right.
      \label{eqtwin}
  \end{equation}

  \noindent where $h_i=\frac{1}{2}\left[ 1+\cos \left( \frac{\pi r_i}{r_0} \right)   \right]$ for $i=1,2$.

  To compare with other schemes, we set this test in $60 \times 60 \times 6$ meshes and with 600 time steps. The result is shown in Table \ref{errortwin}, the error by our scheme is slightly larger than those by CSLAM and SLDG scheme.

  \begin{table}[htbp]
    \caption{Comparison norm errors of deformation flow test of twin cosine bells with other schemes.} \label{errortwin}
    \begin{tabularx}{\textwidth}{l@{\extracolsep{\fill}}ccccc}
    \hline
    Scheme & $l_1$ & $l_2$ & $l_{\infty}$ \\ 
    \hline
    CSLR1 &	0.0722 &	0.1777 &	0.2664 \\
    SLDG P3 &	0.0393 &	0.0673 &	0.1109 \\
    CSLAM &	0.0533 &	0.1088 &	0.1421\\
    \hline
    \end{tabularx}
  \end{table}

  (c) Deformation of correlated cosine bells

  To check the ability of preserve nonlinear correlated relations between two tracers, we used two kinds of tracers. One is the twin cosine bells defined by Eq.\eqref{eqtwin}, and the other one is the correlated cosine bells:

  \begin{equation}
    q^{\text {ccb} }=\Psi (q^{ \text {cb} })
  \end{equation}

  \noindent where $\Psi(q)=-0.8q^2+0.9$.

\begin{figure}[htbp]
  \begin{center}
  \includegraphics[width=0.6\textwidth]{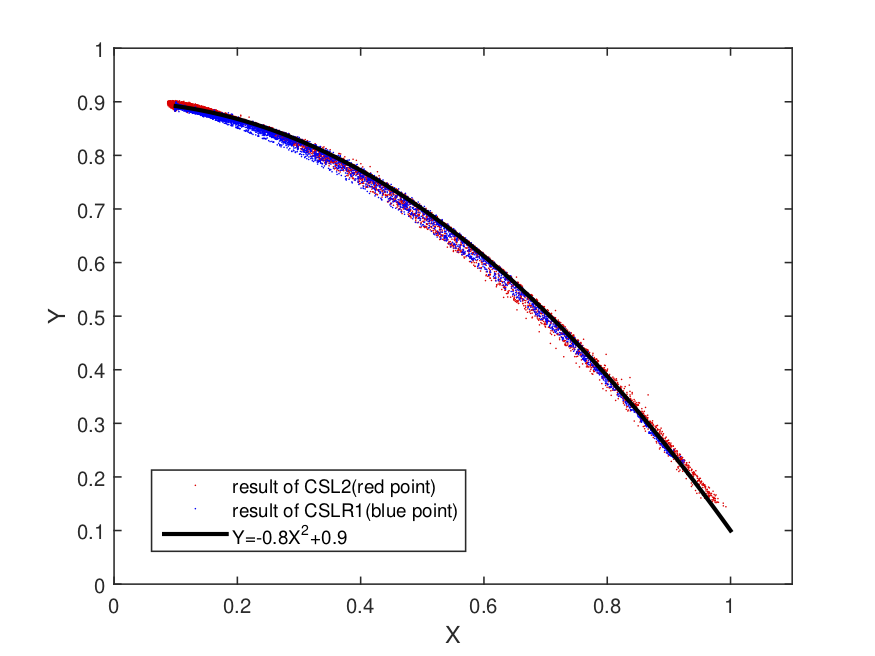}
  \end{center}
  \caption{Scatter plot of nonlinearly correlated cosine bell at $t=T/2$.}\label{scatterplot}
  \end{figure}

This test is conducted on $90\times 90 \times 6$ meshes with 1800 time-steps. The scatter plot of numerical result at $t=T/2$ is shown in Fig. \ref{scatterplot}. The solution of cosine bells is in the X-direction, and the correlated cosine bells is in the Y-direction. The result shows that CSL2 scheme has visible undershoots, while CSLR1 scheme can effectively remove these unphysical undershoots. Table \ref{mixingerrortwin} lists mixing diagnostics. CSLR1 scheme produce more real mixing diagnostic and less range-preserving unmixing. Because of these two schemes are not shape-preserving, some overshooting appears as expected. But the CSLR1 scheme has less overshooting which is consist with the scatter plot.

  \begin{table}[htbp]
    \caption{Mixing diagnostics $l_r$, $l_u$ and $l_o$ at $t=T/2$ for different schemes.} \label{mixingerrortwin}
    \begin{tabularx}{\textwidth}{l@{\extracolsep{\fill}}ccccc}
    \hline
    Scheme & $l_1$ & $l_2$ & $l_{\infty}$ \\ 
    \hline
    CSLR1 &	1.09e-3 &	2.21e-5 &	5.46e-4 \\
    CSL2 &	7.69e-4 &	1.79e-4 &	8.67e-4 \\
    \hline
    \end{tabularx}
  \end{table}

 \section{Summary}
  In this paper, a non-negative and conservative semi-Lagrangian transport scheme based on multi-moment concept has been developed on the cubed-sphere grid. By using two kind of moments such as point value and volume integrated average, a rational function is constructed as spatial approximation function within a single cell. To get a non-negative scheme, two kinds of modifications are conducted on the original CSLR1 scheme. The benchmark tests above demonstrate that CSLR1 scheme is non-oscillatory and can preserve the non-linear correlations between tracers. In case of valley of the transported field, CSLR1 scheme is unable to preserve positivity due to the rational function properity. The definite positivity can be achieved by the easy and effective modifications. In addition, the semi-Lagrangian approach allows the proposed scheme to use large time step, which can greatly improve computational efficiency. Unlike many other non-oscillatory scheme, for example, the WENO limiter used in DG and finite volume method, the proposed model in this paper uses only one cell as stencil for spatial reconstruction, which makes this model very efficient and easy for practical application.

\color{black}

\bibliographystyle{elsarticle-num}
\bibliography{<your-bib-database>}



\end{document}